\documentclass[%
 reprint,
nobibnotes,
 amsmath,amssymb,
 aps,
prb,
]{revtex4-2}

\usepackage{graphicx}
\usepackage{dcolumn}
\usepackage{bm}

\begin{document}


\title{Microscopic processes during ultra-fast laser generation of Frenkel defects in diamond}

\author{Benjamin Griffiths$^{1,2}$, Andrew Kirkpatrick$^{1,2}$, Shannon S. Nicley$^{1,3}$, Rajesh L. Patel$^4$, Joanna M. Zajac$^1$, Gavin W. Morley$^4$, Martin J. Booth$^2$, Patrick S. Salter$^2$, Jason M. Smith$^1$}
\affiliation{$^1$Department of Materials, University of Oxford, Oxford, OX1 3PH, United Kingdom\\
$^2$Department of Engineering Sciences, University of Oxford, Oxford, OX1 3PJ, United Kingdom\\
$^3$Department of Electrical and Computer Engineering, Michigan State University, East Lansing, MI 48824, United States\\
$^4$Department of Physics, University of Warwick, Coventry, CV4 7AL, United Kingdom \\
benjamin.griffiths@materials.ox.ac.uk}

\date{\today}

\begin{abstract}
Engineering single atomic defects into wide bandgap materials has become an attractive field in recent years due to emerging applications such as solid-state quantum bits and sensors. The simplest atomic-scale defect is the lattice vacancy which is often a constituent part of more complex defects such as the nitrogen-vacancy (NV) centre in diamond, therefore an understanding of the formation mechanisms and precision engineering of vacancies is desirable. We present a theoretical and experimental study into the ultra-fast laser generation of vacancy-interstitial pairs (Frenkel defects) in diamond. The process is described by a set of coupled rate equations of the pulsed laser interaction with the material and of the non-equilibrium dynamics of charge carriers during and in the wake of the pulse. We find that a model for Frenkel defect generation via the recombination of a bound biexciton as the electron plasma cools provides good agreement with experimental data, reproducing an effective non-linearity of $\sim$ 40 for Frenkel defect generation with respect to laser pulse energy.
\end{abstract}

\maketitle


\section{Introduction}

Point defects in crystalline semiconductors and insulators are leading candidates for solid state qubits in quantum technologies \cite{Rong} \cite{Simon} \cite{Jelezko}, the atomic-like energy structure of levels deep within the band gap providing long-lived coherent states \cite{Bala}. Colour centres are a class of point defects that display strong optical transitions, which can allow qubit initialisation and readout, and coherent coupling to an optical network \cite{Dutt} \cite{Pezz}. The controlled generation and manipulation of such defects is therefore of great interest. 
Many colour centres take the form X-vacancy, where X is a dopant atom, therefore understanding processes by which vacancies can be generated on-demand is paramount to the development of these technologies. Traditional engineering techniques such as ion implantation \cite{Alekseev} and electron irradiation \cite{Kiflawi} are successful at producing clusters of vacancies, however, often have limited control over the spatial distribution, particularly in the axial direction. This converts into a stochastic yield of colour centres with moderate positional accuracy following an annealing step. Direct laser writing with single ultra-fast pulses has recently been shown as a promising way to excite isolated Frenkel defects (bound vacancy-interstitial pairs) in diamond at desired locations, such that subsequent annealing can produce high quality nitrogen-vacancy centres \cite{Chen}\cite{Sotillo}\cite{Stephen}\cite{Yurgens}. Similar studies in silicon carbide and gallium nitride reveal that laser writing has the ability to produce a host of different colour centres in wide-bandgap materials \cite{Chen3}\cite{Castelletto}. Evidence of a lower intensity regime has recently emerged through multi-pulse processing, in which the laser pulse delivers sufficient energy to mobilise the diffusion of vacancies without further Frenkel defect generation or graphitisation \cite{Kurita} \cite{Chen2} \cite{Kurita2}. These observations open the question of the degree to which point defects can be engineered using ultra-fast laser techniques, and motivate further study. 

The physics of ultra-fast optical pulses interacting with diamond has been studied extensively, however most work has focused on higher energy regimes in which dielectric breakdown occurs and the diamond is graphitised \cite{Simmonds} \cite{Kono} \cite{Hadden}. The breakdown process generates carrier densities of a significant proportion of the atomic density, causing a significant shift in the potential energy surface such that a non-thermal melting occurs through a Coulomb explosion on timescales of $\sim$ 10-100 fs \cite{Medvedev} \cite{Jeschke}. Re-solidification of this material leads to the formation sp$^2$ carbon and voids frozen into the crystal. In addition, longer duration high energy pulses allow for collisions of high energy carriers with lattice ions, which facilitates a thermal melting followed by re-solidification as graphitic material \cite{CWang}. Surface processing in this regime leads to ablation \cite{Kononenko}.

At the lower intensity regimes encountered in \cite{Chen}, very different dynamics are expected. The laser pulse generates non-equilibrium charge carriers through a combination of multi-photon absorption, Zener breakdown and avalanche processes, the relative strengths of which are highly non-linear with respect to the focal intensity  \cite{Keldysh} \cite{Kennedy} \cite{Apostolova} \cite{Gatass}. Following the pulse, carriers thermalise rapidly through carrier-carrier scattering to a temperature much higher than the lattice \cite{Keldysh2}. Scattering with optical and acoustic phonons follows on tens of picosecond timescales such that the excess energy is dissipated into the lattice and the two approach a quasi-equilibrium temperature which decays away through thermal diffusion over several microseconds \cite{Kaiser} \cite{Plamann} \cite{Mao}. The non-equilibrium nature of this system allows for some unusual states of matter including dense excitonic states and at a low enough temperature, condensation into an electron-hole liquid \cite{Keldysh2}. Such states are long lasting and eventually decay through recombination processes.

In these lower energy regimes, the process of defect generation in diamond has yet to be identified. The localisation of energy from carriers to a single atomic site has been predicted via the self trapping of valence biexcitons, which apply a deformation potential of 1.7 eV to a single atomic site and can lead to the breaking of a carbon-carbon bond \cite{Mauri}. However no clear mechanism for Frenkel defect creation has yet been identified.
Density functional theory calculations reveal a formation energy of 7.14 eV for the neutral vacancy in diamond \cite{Deak}, and that an additional energy barrier of at least 0.6 eV exists between the perfect lattice and the Frenkel defect \cite{Salustro}, suggesting that about 8 eV is required to be delivered to a single carbon atom for a Frenkel defect to be created.

The high degree of non-linearity of defect creation with laser pulse energy is a striking feature. To date this non-linearity has been associated solely with the initial photoionisation process and the subsequent energy relaxation leading to defect creation has been assumed to be linear \cite{Barbiero} \cite{Lagomarsino}. In the high energy regime, where lattice damage and breakdown is generally proportional to the total carrier density, these models are able to explain experimental data, but in the lower energy regime lattice damage is not proportional to electron density such that more complex interactions become important. We note that others have created models of some of these interactions, often focusing upon the conduction band electron dynamics, however these models are generally simplified so that the equations can be solved analytically such as ignoring the spatial dependence of carriers and energy transfer to the lattice \cite{Kaiser} \cite{Guenther}. Furthermore many consider the multi-photon processes as a simple power-law relationship with regards to the focal intensity, not considering the band structure or non-linear propagation of the light through the material. Lagomarsino et al showed that degeneracy between valance bands leads to a non-perturbative enhancement of photoionisation leading to higher non-linearities than would be expected in the generation of free carriers \cite{Lagomarsino}. We find that spatial dependence is key as the high electron densities generated lead to metallic behaviour such that non-linear focusing effects and beam attenuation become important to consider. 

Here we seek to develop a quantitative understanding of Frenkel defect generation in diamond by comparing the results of a coupled rate equation model with experiments performed using laser processing. We show that a model in which defect generation occurs via the non-radiative recombination of self-trapped biexcitons combined with thermal excitation over a potential barrier of about 0.5 eV predicts trends that are in good agreement with experiment. Developing understanding of these interactions is key to optimising the processing for the development of quantum technologies.

\section{Experiments}

The key metric used in this paper is the number of Frenkel defects generated by a single laser pulse focused into a diamond sample. Vacancies with a neutral charge state in diamond fluoresce under green illumination at a wavelength of 740 nm known as GR1 fluorescence \cite{Clark}. The intensity of this fluorescence under a given set of excitation conditions therefore provides a measure of the number of vacancies present which can be compared with theoretical predictions. 

The diamond sample used in this study is a type 1b single-crystal supplied by Element Six with 2 parts per billion (ppb) nitrogen concentration. Vacancies were generated using single pulses from a 790 nm, Ti:Sapphire laser (Spectra Physics Solstice) amplified to pulse energies of 10-20 nJ using a chirped pulse amplifier (CPA) and focused into the diamond at a depth of 20 $\mu$m using an oil immersion objective lens (Olympus PlanApo, NA = 1.4). The optical path can be seen in figure S1. Arrays of sites were processed in which the pulse energy was varied by rotating a half-wave-plate before a polariser and measured at the back focal plane of the objective lens. The beam was reflected off a liquid crystal phase-only spatial light modulator (SLM), (Hamamatsu X10468-02), to adjust for aberrations in focusing and thereby minimise the focal volume of the pulse \cite{Simmonds}. Movement between sites in an array was achieved by sample translation on a precision stage (Aerotech air-bearing XYZ) to ensure that the focusing condition remained fixed. Each row of the laser written arrays had 10 sites at 2 $\mu$m spacing of the same pulse energy to allow averaging and the range of pulse energies was chosen with a maximum of significant graphitisation and a minimum where two thousand pulses produced no visible damage. 

Where control of the numerical aperture was required this was achieved by using a blazed grating on the SLM to control the fill factor of the lens. This technique allowed the NA to be varied between 0.95 and 1.4 without changing the objective lens. The laser pulse duration was varied between 120 fs and 1 ps by tuning the compression of the pulse as it passes through the CPA and measured prior to the objective lens via sampling of many pulses by an APE pulse check auto-correlator.  

Photoluminescence measurements were carried out with an excitation wavelength of 532 nm at a continuous-wave power of 1.1 mW and fluorescence was collected between 590 nm and 900 nm in a home-built confocal microscope. To determine the fluorescence intensity at a given laser power, 2D Gaussian peaks were fitted to the confocal images of laser written sites giving a value for the brightness from the integrated intensity. The background fluorescence was measured 1 $\mu$m from the laser written points which was then subtracted from the intensity value such that the measured intensity value came solely from the laser process. The recorded fluorescence intensity was determined by averaging the measured values from ten sites.  

Figure 1a shows an example of a fluorescence image of an array of laser written sites. Spectral analysis (figure 1b) and wide-field transmission imaging reveal that up to 18 nJ the fluorescence is dominated by the GR1 peak at 740 nm, whereas at higher pulse energies a significant B-band peak is observed showing the onset of graphitisation. The spectrally integrated fluorescence intensity as a function of write-pulse energy is shown in blue on a logarithmic plot in figure 1c. The intensity increases by a factor of 50 between 16 and 18 nJ giving an average non-linearity of around 40, but then quickly saturates and is approximately linear at pulse energies above 20 nJ. Overlaid in black is the best fit simulated data from numerically solving the partial differential equation (PDE) model developed in the following section.

\begin{figure}
  \centering
  \includegraphics[width=8cm]{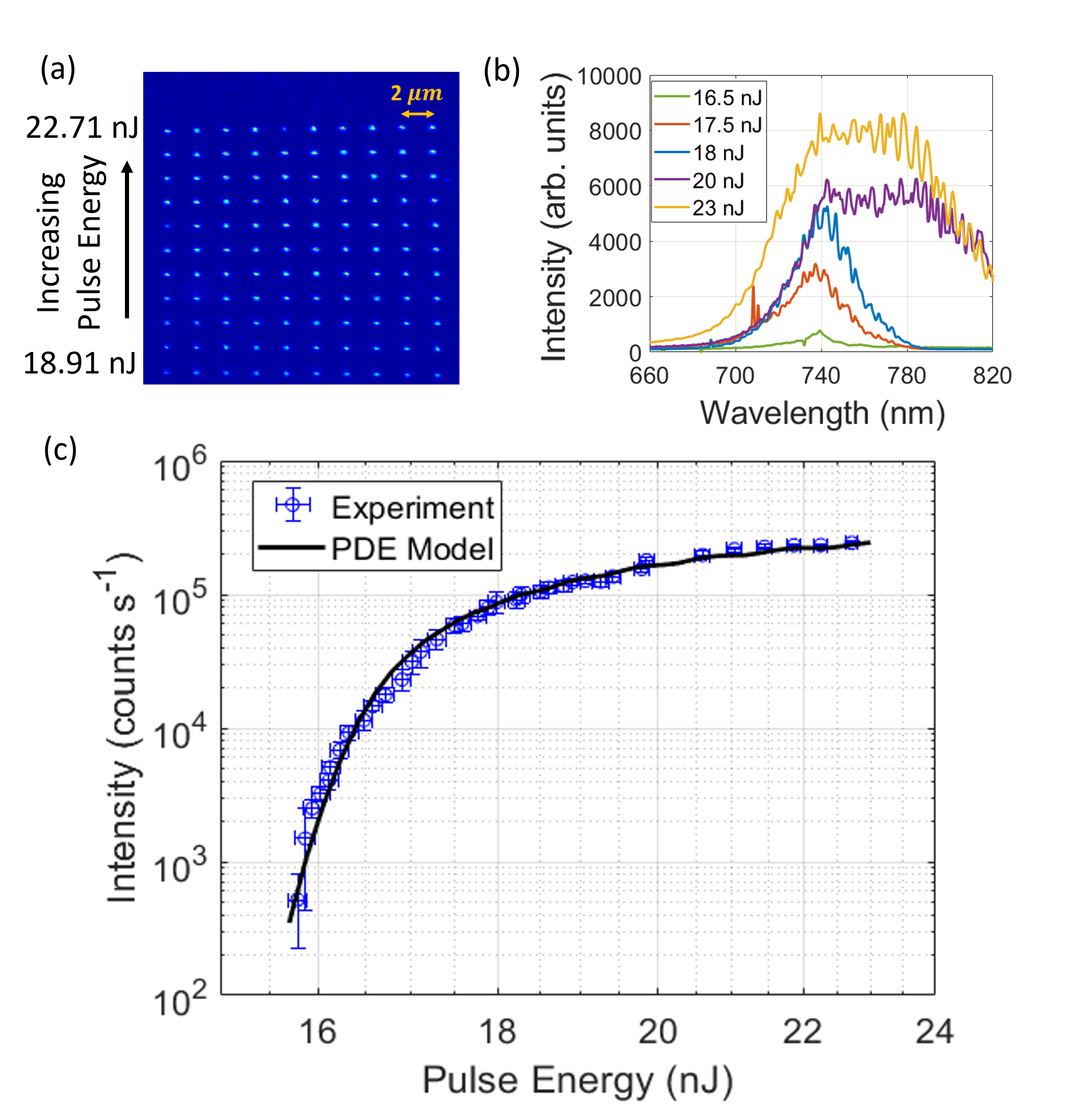}
   \caption{a. A confocal image of the highest energy sites in the array. b. Spectral composition of laser written sites, fluorescence is centred at a peak at 740 nm corresponding to lattice vacancies which broadens and shifts to longer wavelengths for higher pulse energies indicating further damage and graphitisation. c. The average fluorescence intensity measured from single pulses against pulse energy (blue) overlaid with simulated data (black).}
  \label{fig:fig1}
\end{figure}

\section{Rate Equation Model}

We model the dynamics of the laser pulse interaction with the diamond using a set of coupled non-linear differential equations and solve numerically using finite difference methods describing the evolution of the system from the onset of the pulse to 100 ps after the pulse passes through the focus. This allows sufficient time for all defect generation to occur, although the lattice temperature remains elevated and the heat generated takes a further nanosecond to dissipate. The rate equations are applied over the two spatial coordinates corresponding to the cylindrically symmetric focal spot, covering an axial range of 3 $\mu$m about the focal plane and a radial distance of 200 nm, which was found to be large enough that edge effects are negligible. A time-step of 0.1 fs was chosen, a pixel size of 17 nm in radius and 38 nm in depth with an intraband energy resolution of 0.02 eV allowing for a reasonably fine mesh of the region of interest.

We used eight rate equations to model the evolution of the following parameters: (1) optical intensity, (2 and 3) electron and hole concentrations, (4) exciton concentration, (5 and 6) free and self-trapped biexciton concentration, (7) lattice temperature, and (8) Frenkel defect concentration. Carrier concentrations are expressed as distributions over energy up to a maximum of 25 eV. Each rate equation takes the form of a Boltzmann equation \cite{Boer}, modeling the change to the distribution as a sum of all processes that act upon that distribution in phase space. Full details are given in the supplementary information - here we describe the key features.
 
The optical intensity and electric field as functions of position and time are calculated based on the focal intensity distribution of the focused laser pulse as it propagates through the material, accounting for absorption of photons by electrons within the material and the changes to how the pulse propagates due to the generated electron plasma including self-focusing and filamentation \cite{Mao} \cite{Kolesik}. For the input laser pulse we assume a top-hat intensity distribution with a perfectly aberration-corrected phase distribution at the back focal plane of the objective lens and use standard Fourier optics to construct the 3D distribution of the laser intensity profile \cite{Hecht}\cite{Jesacher}. The temporal evolution is modelled as a Gaussian envelope to the sinusoid at the laser frequency. 

The rate equations for free carriers include all relevant generation, recombination, scattering processes and exciton formation (see supplementary information). Generation rates were calculated using the band structure of diamond obtained using density functional theory and numerical integrals were evaluated using the Euler finite difference method. For the interaction rates we use the collision time approximation, drawing values for average interaction times from previous works (table I). This approach allows carrier scattering rates to be evaluated as explicit functions of carrier energy, but averaged over momentum space. This is a reasonable approximation since carrier scattering homogenises the momentum distribution across the Brillouin zone after a few scattering events.

\begin{table}
\begin{ruledtabular}
\begin{tabular}{lcc}
Parameter & value & ref.\\
\hline
Electron-electron scattering & Eq. S4 & \cite{Kozak}\\
Electron-phonon scattering & Eq. S7 & \cite{Tandon}\\
Electron-phonon-photon scattering & Eq. S11 & \cite{Kennedy} \\
Radiative electron-hole recombination & 2.3 $\mu s$ & \cite{Lipatov}\\
Radiative exciton recombination & 350 ns & \cite{Morimoto}\\
Radiative biexciton recombination & 7.3 ns & \cite{Omachi} \\
Radiative self-trapped biexciton recombination & 7 ns & \cite{Mauri} \\
Optical phonon energy & 0.15 eV & \cite{Monserrat}\\
Acoustic phonon energy & 0.08 eV & \cite{Monserrat}\\
Exciton binding energy & 80 meV & \cite{Clark2} \\
Biexciton binding energy & 12 meV & \cite{Katow} \\ 
Biexciton self-trap deformation potential & 1.74 eV & \cite{Mauri} \\
\end{tabular}\\
\label{tab_lifetimes}
\normalsize
\vspace{0.5cm}
Table I: Interaction parameters and sources
\end{ruledtabular}
\end{table}
\normalsize

The lattice temperature rate equation has two terms corresponding to lattice heating through carrier-phonon scattering and heat diffusion. We define a graphitisation threshold when the carrier density reaches 9\% of the atomic density \cite{Bennington} \cite{Sampfli} or as when the lattice exceeds the melting temperature (4300 K). An electron temperature was calculated through least squares fitting of an exponential function to the electron distribution.

The rate of formation of excitons was calculated using the local concentration of electron-hole pairs with kinetic energy less than the excitonic binding energy \cite{Mao} \cite{Selbmann}, which in diamond is $\mathcal{E}_{X} = 80$ meV \cite{Clark2}. The interaction timescale depends upon the Debye length of carriers, which in turn is dependent on values such as the electron temperature and concentration. This timescale is of the same order as the carrier-carrier scattering as both take place through the screened Coulomb interaction so we utilise the same empirical parameter for both processes. Exciton-carrier scattering, exciton-phonon scattering, and recombination processes are also included. Finally the formation and dissociation of biexcitons contributes to the exciton population.

In a similar manner to exciton formation, the rate of biexcitons formation is calculated from the concentration of pairs of excitons with kinetic energy lower than the biexcitonic binding energy of $\mathcal{E}_{bX}=12$ meV \cite{Katow}. As with the excitons, scattering with other carriers and with phonons leads to thermalisation of the biexciton population. The rate of radiative recombination of the biexcitons is determined using parameters from \cite{Omachi}. biexcitons are more localised than than single excitons and can self-trap, exerting a deformation potential of $\mathcal{E}_{DP}=$1.74 eV to the lattice which can lead to the breaking of a carbon-carbon bond \cite{Mauri}. 

\begin{figure}
  \centering
  \includegraphics[width=7cm]{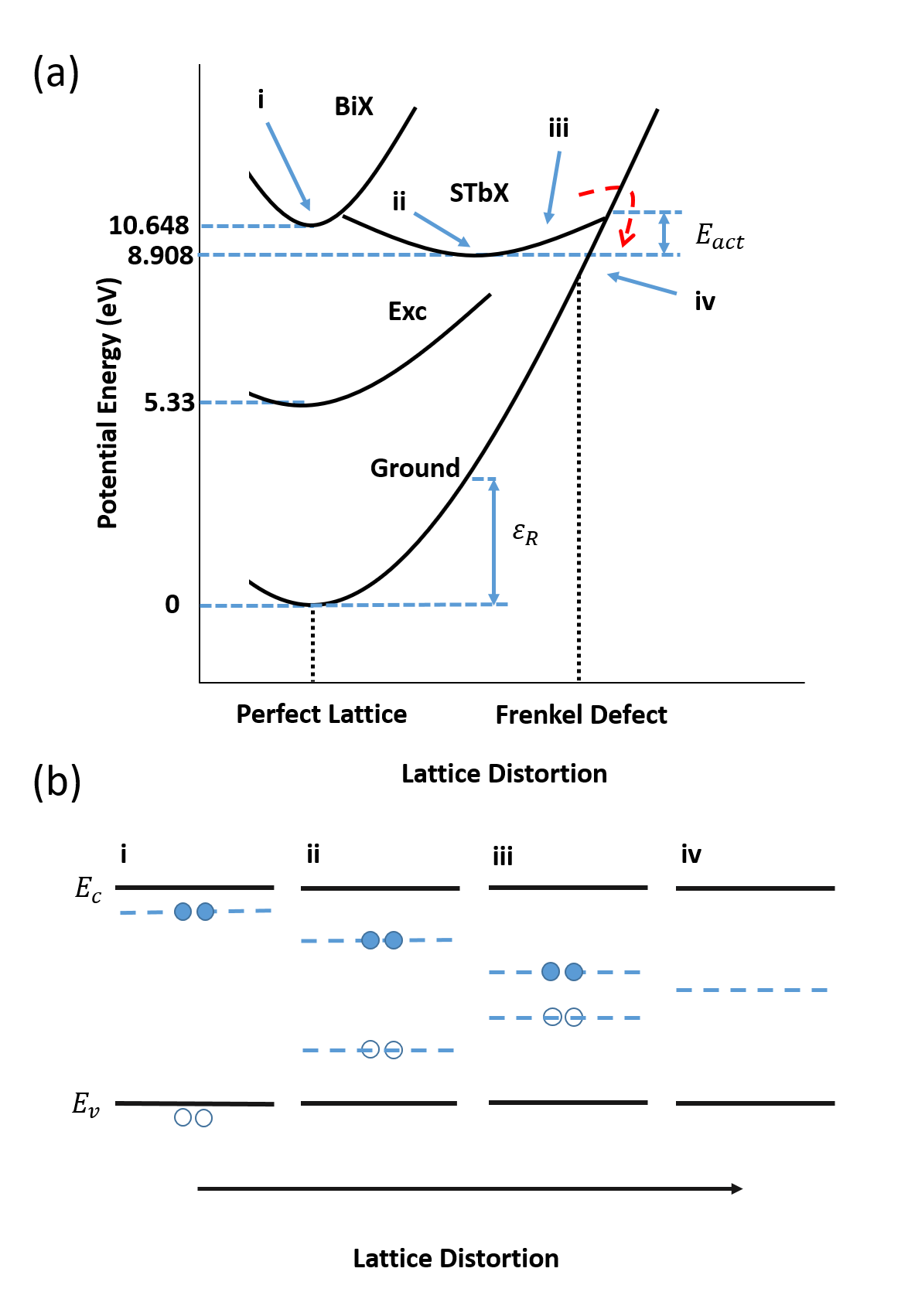}
    \caption{a. Schematic of the adiabatic potential energy curves of the ground state, free biexciton state (BiX), free exciton state (Exc), and the self-trapped biexciton state (STbX) against lattice distortion. The red dashed line shows possible non-radiative decay processes of the STbX's and the blue dashed lines indicate the activation energy $E_{act}$ and relaxation energy, $\epsilon_R$ of the ground state. b. A schematic of the proposed band structure during the non-radiative recombination of the self-trapped biexciton. The bonding, anti-bonding states in the band-gap approach each other as the lattice is further deformed allowing non-radiative recombination.}
  \label{fig:fig2}
\end{figure}

Mauri \emph{et al.} proposed that following self-trapping, radiative recombination of the biexciton would follow with a large Stokes shift of 3.23 eV on timescales of $\tau_r \sim$ 7 ns leaving a single exciton and the lattice partially distorted, leading to a localised graphitisation \cite{Mauri}. Here we propose an alternative mechanism, a non-radiative recombination pathway, transferring the energy of the entire biexciton to the lattice and forming a Frenkel defect. The energetics of this proposed process are depicted in figure \ref{fig:fig2}. The self-trapped biexciton (STbX) localised on the broken bond has a minimum energy of 8.9 eV which makes the formation of a 7.14 eV Frenkel defect energetically favourable. The driving force behind the process is depicted in figure \ref{fig:fig2}b. The STbX takes the form of two deep levels within the band-gap, a doubly occupied anti-bonding level 1.7 eV below the conduction band edge and an empty bonding level 1.6 eV above the valence band maximum, both localised on the broken bond \cite{Mauri}. It is therefore possible that further deformation of the lattice pushes the deep levels towards mid-gap as shown in figure \ref{fig:fig2}b allowing recombination through electron capture and creating the Frenkel defect.

Similar processes are known to occur in other materials. Bang \emph{et al.} showed that 2 electron-hole pairs within InGaN can non-radiatively recombine via the formation and recombination of a Frenkel pair \cite{Bang} and self trapped excitonic states have been shown to lead to the permanent formation of Frenkel defects in SiO\textsubscript{2} and other dielectrics \cite{Mao} \cite{Itoh} \cite{Williams}. In both cases, the deformation of the lattice leads to the formation of states within the band-gap that allow energy transfer from the electronic system to the lattice. 

The rate of Frenkel defect formation is calculated from the STbX population, with a rate constant determined by a  thermally activated electron capture process \cite{Shenk}.

\begin{equation}
\tau_{FD}(r,z,t,\mathcal{E})=\frac{h}{\mathcal{E}_{ph}}\exp\left(\frac{\mathcal{E}_b}{k_BT_{L}}\right)
\end{equation}
\normalsize

where $\mathcal{E}_{ph}$ is the optical phonon energy, $\mathcal{E}_b$ is the energy barrier \cite{Bang} and $T_L$ is the lattice temperature. The height of this energy barrier is established by assuming that the shape of the self trapped biexciton potential energy surface is symmetric about its minimum and that the formation energy is equal to the activation energy minus a quarter of the relaxation energy of the lattice  which holds in the high temperature limit \cite{Shenk}. For the STbX, this suggests an energy barrier of 1.74 - $\frac{4.85}{4}$ = 0.53 eV. This value is consistent with our experimental data and is discussed further in section 5. Using this value of $\mathcal{E}_b$ in equation (1) results in a lifetime for Frenkel defect formation that is shorter than the radiative lifetime of the STbX for lattice temperatures above about 450K.

We neglect any processes that are facilitated by the creation of the Frenkel defects. This is a reasonable assumption since defect formation does not begin until several picoseconds after the optical pulse has passed. We also ignore any recombination of Frenkel defects, although the recombination of a fixed fraction of the defects would appear as a scaling factor and would otherwise not alter the results of the model.

\section{General Characteristics}

\begin{figure}
\centering
\includegraphics[width=8cm]{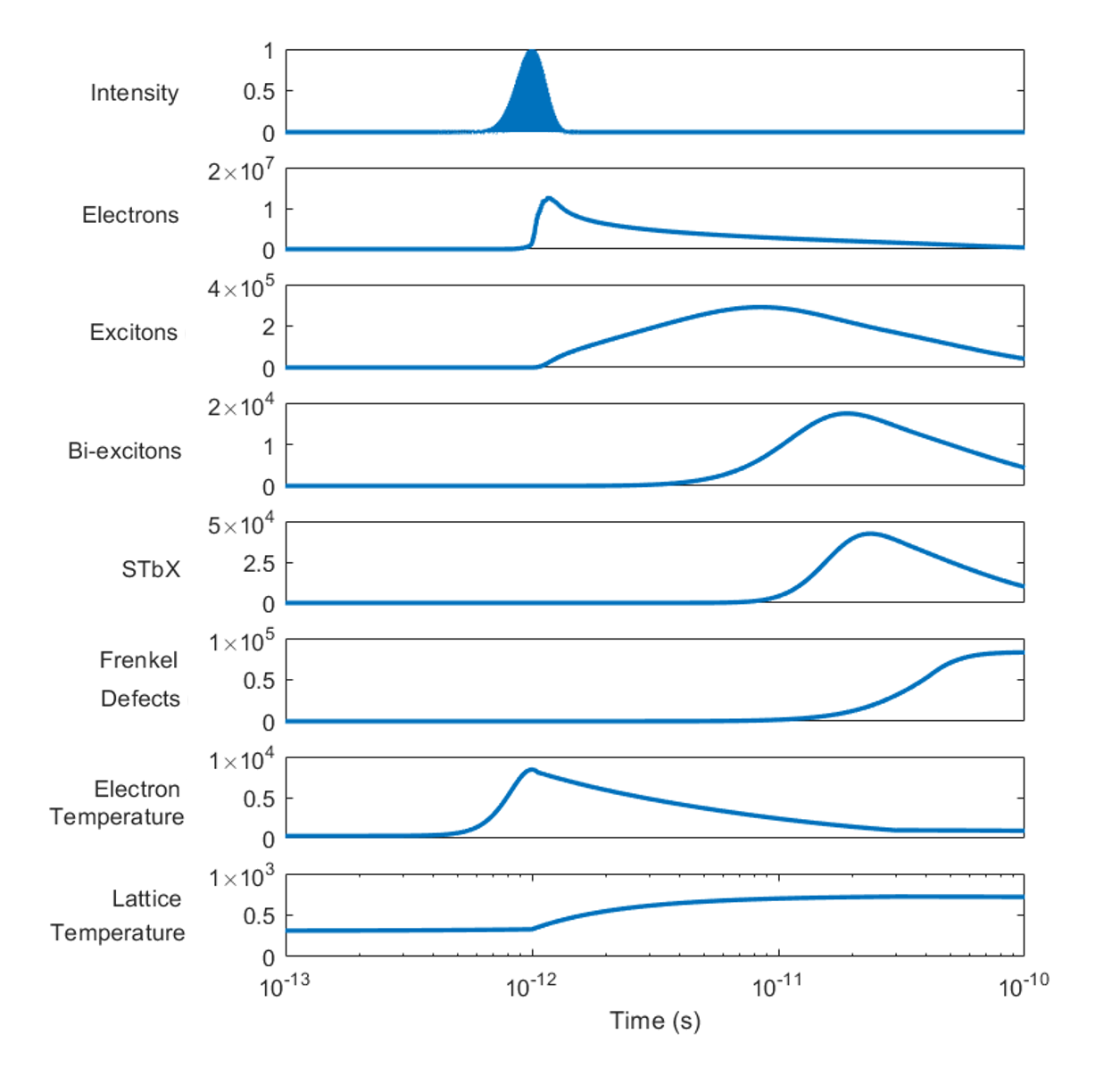}
\caption{Time dependence of parameters simulated in the PDE model. The peak of the laser pulse is set to 1 ps to facilitate a logarithmic time scale capturing the range of dynamics. Through charge neutrality, the number of holes is equal to the number of electrons.}
\label{fig:fig3}
\end{figure}

Figure \ref{fig:fig3} displays the simulated time dynamics for the various parameters generated by an 6 nJ, 300 fs pulse at the focus of the laser pulse, allowing the relative timings of the processes to be observed. Electron generation occurs quickly during and following the pulse, initially via photoionisation but the majority being generated later via cascade processes. The electron concentration peaks around 0.25 ps after the peak of the pulse intensity and then decays multi-exponentially due to non-radiative recombination processes such as Auger processes and excitonic formation. Rapid carrier-carrier scattering ensures that the different carrier types are at thermal equilibrium with each other and re-ionises the majority of any excitonic states that form. Immediately after the pulse the carrier temperature reaches several thousand degrees while lattice temperature remains low. Electron-phonon scattering then causes the carrier temperature to decay rapidly and the lattice temperature to increase until equilibrium is reached after around 30 ps. 

The cooling of the hot carriers also triggers the increase of exciton and then biexciton populations, with peak populations at times of about 9 ps and 20 ps after the laser pulse. The STbX population and the resulting population of Frenkel defects then follow some 22 ps and 30 ps after the laser pulse has passed respectively. We note that by this time the lattice temperature is sufficiently high that the proposed thermally activated Frenkel defect formation process will be considerably faster than radiative recombination of the STbX discussed by Mauri \emph{et al.}. At $\sim$ 20 ps the electron and lattice temperatures reach a quasi-equilibrium. This temperature will then decay over timescales of order microseconds through thermal diffusion. 

\begin{figure}
  \centering
 \includegraphics[width=8cm]{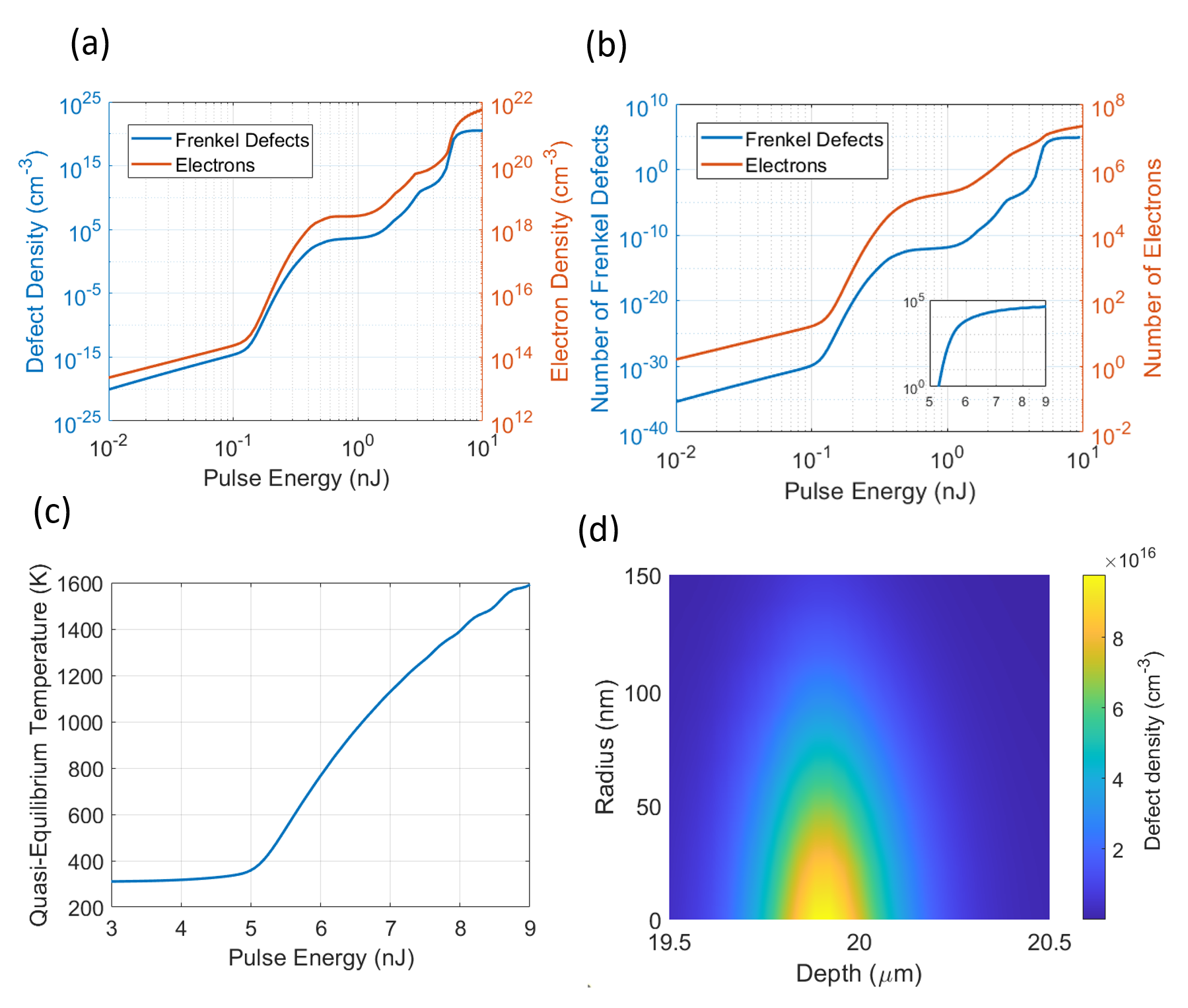}
   \caption{a. Maximum densities of carriers and defects generated against pulse energy. In red the number of electrons that are created against pulse energy and in blue, the simulated number of defects generated. b. Numerical integration of the coupled PDE model over energy, space and time. Inset displays the number of defects generated with probability above 1. c. The quasi-equilibrium temperature reached by the electron-lattice system at the centre of focus against pulse energy. d. Simulated spatial distribution of defects from a pulse of 5.5 nJ focused 20 $\mu$m into the diamond.}
  \label{fig:fig4}
\end{figure}

Figures \ref{fig:fig4}a and b shows the peak density and total number of photogenerated electrons as a function of write-pulse energy along with the peak density and total number of Frenkel defects generated per pulse. The photoelectron generation curve follows a similar dependency to that from photoionisation alone as shown by Lagomarsino \emph{et al.} \cite{Lagomarsino}, (figure S3 in the supplementary information), however the impact ionisation increases and smooths out the number of electrons generated. At high pulse energies there is an uptick in the non-linearity of carriers at the centre of focus. This is because carrier densities become high enough that the carrier mobility rapidly decreases leading to an accelerated carrier avalanche. This is rapidly saturated as Auger processes reach a quasi-equilibrium with the impact ionisation and plasma effects suppress photoionisation. Away from the centre of the focus, growth continues to occur leading to a smoothing out of the total number of carriers.
For pulse energies below about 4 nJ the defect generation curves in figure \ref{fig:fig4}a and b follow a similar pattern to that of electron generation, but with a non-linearity about four times larger due to the biexciton-induced formation mechanism. The Frenkel defect generation probability passes 1 at $\sim$ 5 nJ with a fast rise in the number of defects generated over a very small range in pulse energies which quickly saturates. This extremely steep dependence of defect formation on pulse energy occurs primarily as a result of  two effects. The reduced carrier mobility not only produces an accelerated carrier avalanche but also leads to a significantly accelerated formation of excitonic states following cooling of the carriers. In addition, the temperature at which the electron-lattice system reaches a quasi-equilibrium for these pulse energies becomes significant, this accelerates the non-radiative recombination of the self trapped biexcitons in competition with radiative processes. The quantum efficiency for Frenkel defect formation, defined as $\eta=\frac{\tau_r}{\tau_r+\tau_{FD}}$ and using equation (1), where $\tau_r$ is the radiative recombination timescale, rises from $\sim$ 10$^{-3}$ at room temperature to near unity at temperatures above 500 K. The peak quasi-equilibrium temperature reached by the carrier-lattice system against pulse energy is shown in figure \ref{fig:fig4}c. At low pulse energies the maximum carrier densities are low enough that the quasi-equilibrium temperature reached is not significantly above the ambient temperature, however for higher pulse energies, it rises rapidly such that Frenkel defect formation will quickly become efficient over radiative recombination processes.

The rapid saturation in Frenkel defect number in figures \ref{fig:fig4}a and b is again attributed to two processes. At these high pulse energies, there is a saturation to the electron density growth which becomes high enough that efficient Auger processes take equal precedence with avalanche ionisation. In addition, the non-radiative recombination of self-trapped biexcitons becomes efficient compared with radiative processes as the electron-lattice equilibrium temperature is high.

The spatial distributions of the generated Frenkel defects at pulse energies of 5.5 nJ and 7 nJ are shown in figure \ref{fig:fig4}e and figure S5. At the lower pulse energy the distribution approximately follows the diffraction limited focal intensity distribution of the microscope scaled by the non-linearity, while at the higher pulse energy a distortion is observed that results from the metal-like properties of the high density of free carriers \cite{Mao} \cite{Kaiser} \cite{Combescot}.  This changes the focusing of the incoming light leading to absorption at shallower depths and attenuation of the beam, altering the spatial distribution of defect formation. The simulated full-width half maximum (FWHM) for the 5.5 nJ pulse is $\sim$ 250 nm in depth and $\sim$ 50 nm in radius. These values are very similar to the measured positional accuracy of laser written NV centers by Chen \textit{et al.} \cite{Chen2}. Whilst this is not a direct comparison since the model does not include the annealing required to form NV centres from the Frenkel defects, the consistency between the two gives further confidence in the simulation results. 

\section{Comparison with Experimental Data}

In the experiments, the onset of visible GR1 fluorescence occurs at a pulse energy of about 16 nJ, almost three times higher than that predicted from the simulation results of figure 4. There are a number of possible reasons for this discrepancy, including unaccounted aberrations in the focal spot, an underestimate of optical losses between the objective lens and the focal spot, and inaccuracies in parameters such as scattering rates and binding energies that will influence the efficiency of biexciton formation. We treat this unknown scaling factor as a fitting parameter. A second scaling factor is needed in order to compare the number of Frenkel defects generated with the measured GR1 fluorescence intensity. This scaling factor is dependent on the fluorescence quantum efficiency of the GR1 defects and on the efficiency of the fluorescence microscope used for the measurements. Respective values of 2.84 and 3.47 were found for these two scaling factors, providing the fit shown in figure 1 with an R$^2$ of 0.9977 and adjusted R$^2$ of 0.9976. These scaling factors were found by first taking the base 10 logarithm of the model and experimental data such that linear regression could be used.

\begin{figure}
  \centering
 \includegraphics[width=8cm]{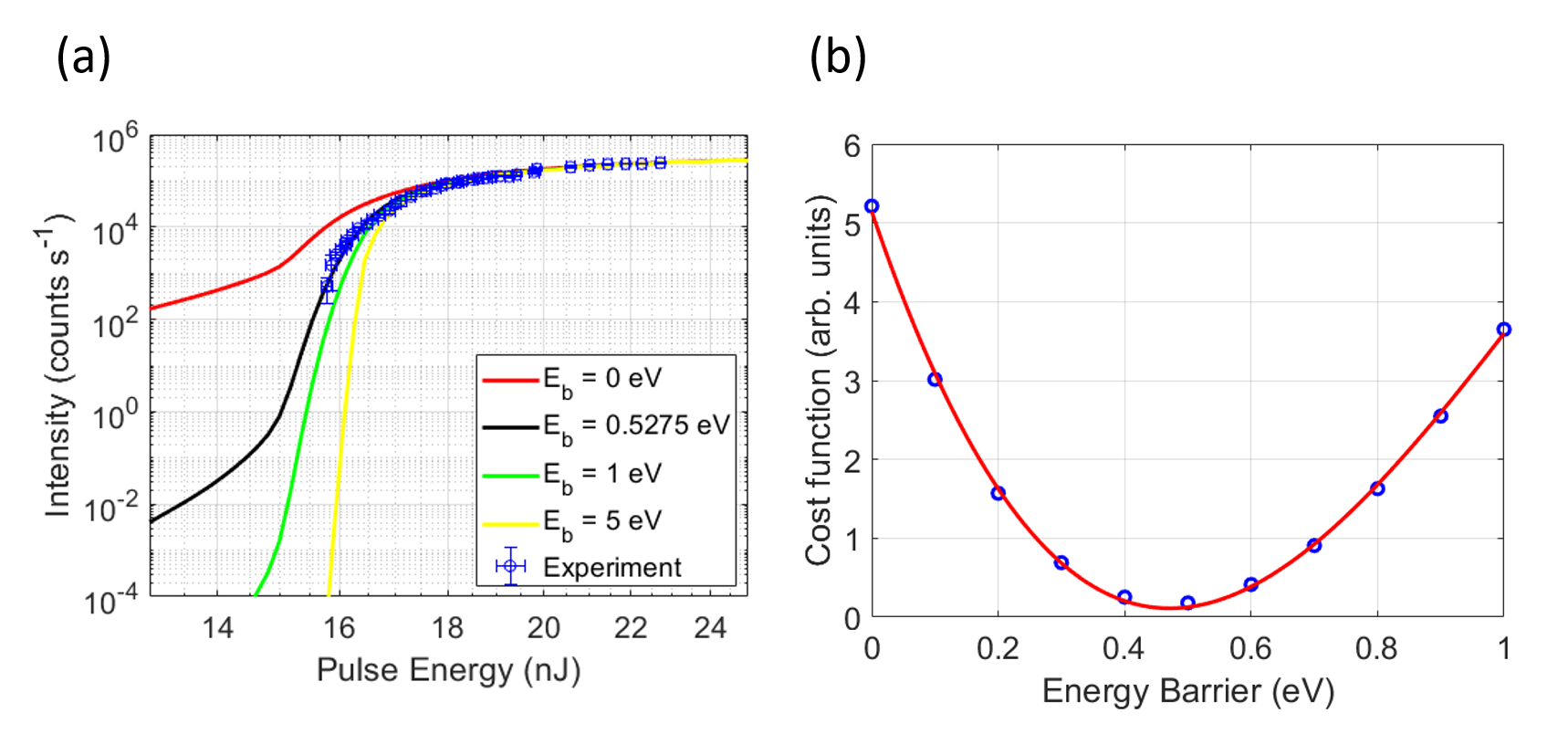}
   \caption{a. Simulated defect vs pulse energy curves for different values of $\mathcal{E}_{b}$ fitted to experimental data through non-linear least squares fitting. b. Minimisation of the cost function for the energy barrier to non-radiative recombination reveals a value of $\mathcal{E}_{b}=$ 0.47 $\pm$ 0.01 eV fits best to the experimental data.}
  \label{fig:fig5}
\end{figure}

Figures \ref{fig:fig5}a and b reveal the sensitivity of the model to the energy barrier height $\mathcal{E}_{b}$. Each simulation curve is scaled on both the $x$ and $y$ axes as described above to provide a best fit to the experimental data. Figure \ref{fig:fig5}a reveals that the magnitude of the energy barrier plays a significant role in determining the magnitude of the non-linearity. The sum of the squared residual distance between the experimental data and the model is plotted versus $\mathcal{E}_{b}$ in figure \ref{fig:fig5}b. A polynomial fit reveals that the residuals are minimised at a value of $\mathcal{E}_{b}=$ 0.47 $\pm$ 0.01 eV with a direct overlay shown on-top of the experimental data shown in figure \ref{fig:fig1}c. The high degree of agreement between the model and experiment gives weight to the hypothesis that most of the energy required to form a Frenkel defect comes from the recombination of self-trapped biexcitonic states. In addition, the need for biexcitonic states explain the localisation of energy to single lattice sites.

\begin{figure*}
  \centering
  \includegraphics[width=\textwidth]{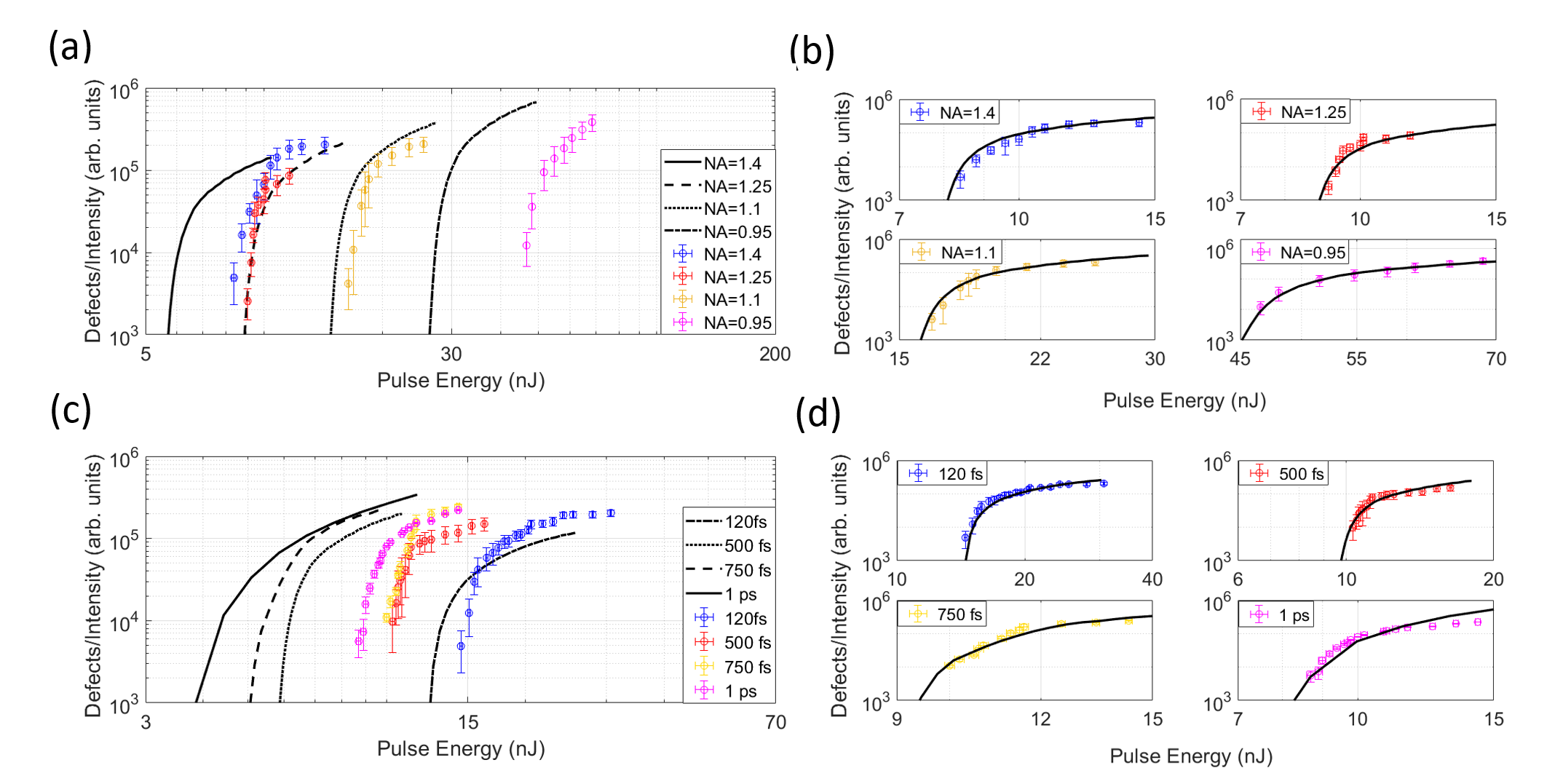}
    \caption{a. Simulated (lines) defect generation probability as a function of pulse energy for differing NA of focus  alongside experimental data (points). The NA's range from 0.95 to 1.4. b. Data and simulation for differing NA of focus overlaid to show agreement in the non-linearity. c. Simulated  (lines) defect generation probability as a function of pulse energy for differing pulse durations alongside experimental data (points). The pulse durations range from 120 fs to 1 ps. d. Data and simulation for differing pulse durations overlaid to compare the non-linearity.}
  \label{fig:fig6}
\end{figure*}

Finally we explore the dependence of Frenkel defect generation on two practical parameters: the numerical aperture (NA) of fabrication laser focus and the duration of the laser pulse. Reducing the NA of the focusing lens reduces the focal intensity and thereby increases the pulse energy needed for defect generation. In addition some modest change in the dynamics might be expected as a result of a reduction in carrier diffusion. We set initial conditions for various focusing NA, ranging between NA of 0.95 and 1.4. The corresponding simulated and experimental data are seen in figure \ref{fig:fig6}a.

The general trend described above is visible in both the simulation results and experimental data. We note that in the experiments, a numerical aperture of 1.4 showed no reduction in required pulse energy compared with NA = 1.25. This could indicate the presence of uncorrected aberrations such that the outermost rays passing through the objective do not contribute measurably to the peak focal intensity. Also the required increase in pulse energy for NA = 0.95 is somewhat larger than predicted. Figure \ref{fig:fig6}b shows the same experimental data with the simulation data-sets independently scaled to provide best fits. The dependence on relative pulse energy does not vary substantially with NA and the simulations fit well to the data throughout.

The effect of varying the pulse duration between 120 fs and 1 ps is shown in figures \ref{fig:fig6}c and d where, as above, panel d includes a rescaling of the individual simulation data-sets to provide best fits to the experimental data. Somewhat counter-intuitively, the general trend in both experimental data and the simulation results is for shorter pulses to require higher pulse energies for defect generation. This is because, although shorter pulses achieve the same focal intensity for lower pulse energy, they generate fewer carriers due to a reduction in free carrier absorption and subsequent impact ionisation \cite{Kurita}. The effect of this change is particularly marked in the results of the simulation for the 120 fs pulse, where the threshold for Frenkel defect generation is a factor of two larger than that for the 500 fs pulse. In the experiment only a 30\% increase in threshold is observed, which may indicate that the pulse envelope for the shortest pulses is less well-approximated by the Gaussian distribution used in the model. Experimentally, the observed non-linearity is again relatively unchanged; in contrast the model predicts a noticeable reduction in non-linearity for the longest pulses.

\section{Conclusion}

In summary, we have studied the generation of Frenkel defects in diamond in response to a single, sub-picosecond laser pulse. Experimental data show a non-linearity of around 40 at the onset of Frenkel defect generation, followed by rapid saturation in the number of Frenkel defects created. A finite-difference time-domain model based on coupled non-linear rate equations describes the carrier excitation and relaxation processes, with Frenkel defect formation attributed to the thermally activated non-radiative relaxation of self-trapped biexcitons. The model shows good agreement with experimental data for the dependence of defect formation on pulse energy, and an activation barrier height of 470 $\pm$ 10 meV is derived from fit. The effects of laser focal spot size and pulse duration were explored, revealing that impact ionisation of hot carriers is an important contributing factor in generating the high carrier densities needed, such that longer pulses, in which carrier heating plays a more significant role, generate defects more efficiently than do shorter pulses. 

The model described here is quite general and could in principle be applied to other transparent materials such as glass or silicon carbide, providing a useful tool for understanding defect generation mechanisms in these materials, or to describe laser annealing processes as reported in \cite{Chen2}. As such it will assist in the optimisation of fabrication parameters for the defect engineering of materials in solid-state quantum devices and lead towards greater understanding of ultra-fast laser interaction with wide-bandgap materials.

\section{Acknowledgments}

This work is supported by the UK Hub in Quantum Computing and Simulation, part of the UK National Quantum Technologies Programme with funding from UKRI EPSRC grant EP/T001062/1 and additional support was received from EPSRC grant (Grant No. EP/R004803/01).

The coupled differential equation model was developed and numerically solved by Benjamin Griffiths. Discussions with Jason Smith, Patrick Salter, Martin Booth, Shannon Nicley, Joanna Zajac and Gavin Morley enabled clarification on various mechanisms. The band-structure was calculated by Andrew Kirkpatrick. Direct laser writing was performed by Benjamin Griffiths and Patrick Salter and fluorescence characterisation was conducted by Benjamin Griffiths, Rajesh Patel and Shannon Nicley. Data analysis and fitting of the model to experiment was conducted by Benjamin Griffiths. The manuscript was written by Benjamin Griffiths and Jason Smith in discussions with Patrick Salter and Joanna Zajac.

\bibliographystyle{unsrt}  


\clearpage
\pagebreak

\section{Supplementary Information}

The following sections will outline the optical apparatus used for laser writing and the coupled partial differential equation model used in this work. The set of coupled equations defining each system are S1, S3, S14, S15, S17, S18 and S19.

\subsection{Experimental Apparatus}

Laser writing was undertaken using a regeneratively amplified Ti:Sapphire laser (Spectra Physics Solstice). A schematic of the laser writing apparatus is shown in Figure S1. The regeneratively amplified laser undergoes further amplification via a chirped pulse amplifier (CPA) to achieve intense ultra-short pulses that are above the breakdown threshold of the laser cavity. The output laser pulses were initially expanded onto a liquid crystal phase-only spatial light modulator (SLM) (Hamamatsu X10468-02) for iterative aberration control, and then imaged in a 4f system onto the back aperture of a $\times$60 1.4NA oil immersion objective lens (Olympus PlanApo). Diamond samples were mounted on precision translation air bearing sample stages (Aerotech x-y: ABL10100; z: ANT95-3-V) which provide movements in all three directions. A LED transmission microscope was built alongside and imaged onto a CCD to monitor the sample during the laser processing. Control of the pulse duration was enabled through tuning the CPA and control of the numerical aperture was enabled through application of a blazed grating to the SLM. Pulse energy was controlled through rotation of a half-wavelength wave-plate and linear polariser and measured at the back focal plane of the objective lens. The spatial light modulator is key to achieving the high intensities required for Frenkel defect generation by applying the reverse phase to focal aberrations introduced during focusing. The schematic in the figure shows the phase pattern applied to correct for spherical aberration introduced by the large mismatch in refractive indices at the oil-diamond interface \cite{Jesacher}. An arbitrary aberration can be corrected for through applying a superposition of orthogonal polynomials on the unit disk \cite{Hecht}.

\renewcommand{\thefigure}{S1}
\begin{figure}
    \centering
    \includegraphics[width=8cm]{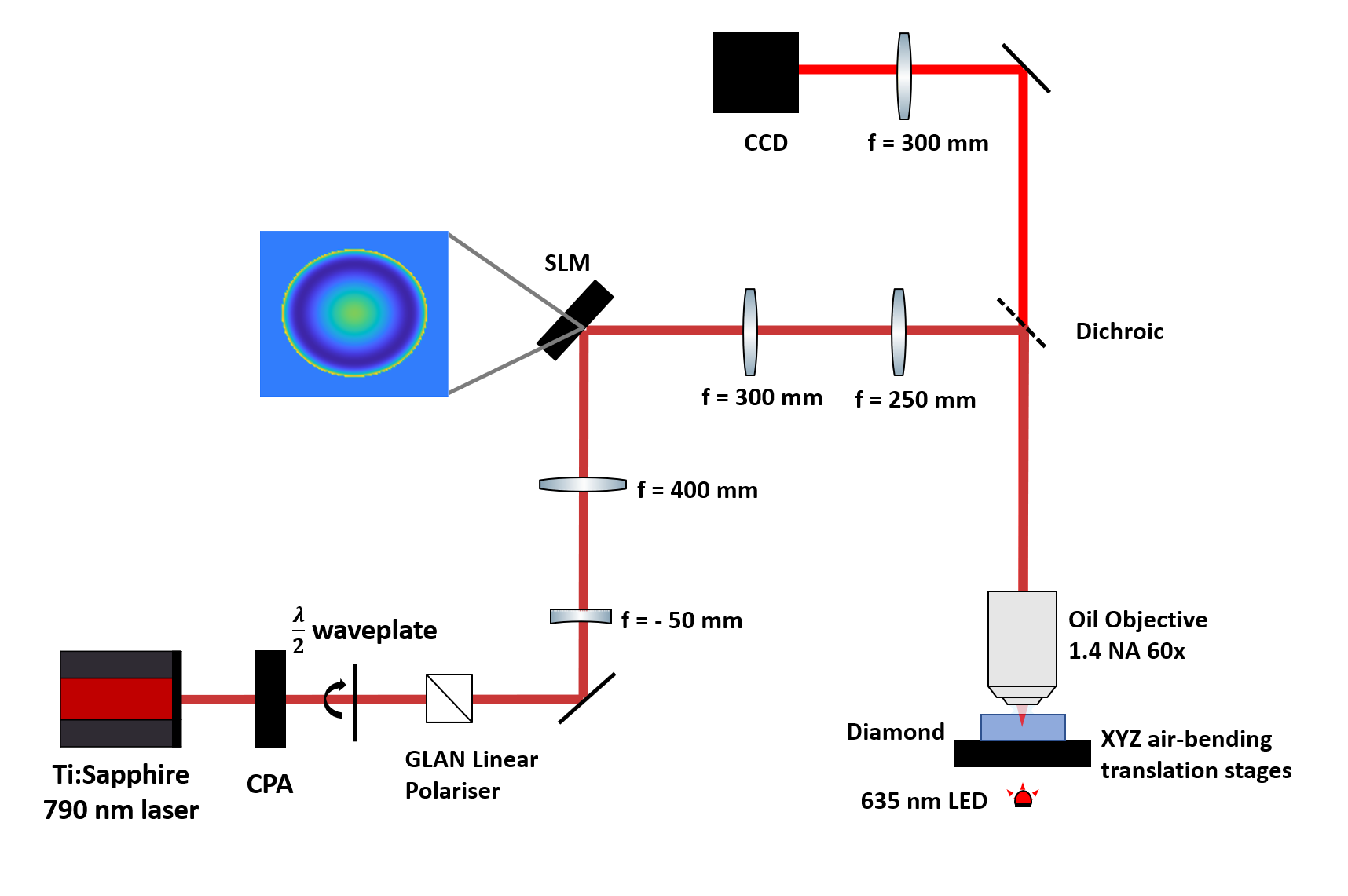}
    \caption{The optical path used to focus the Ti:Sapphire into the diamond for laser writing. Control of the pulse duration was enabled through tuning the CPA and control of the numerical aperture was enabled through application of a blazed grating to the SLM.}
    \label{fig:my_label}
\end{figure}

\subsection{Optical Intensity Distribution}

The optical intensity was calculated according to the following equation.

\begin{equation} \tag{S1}
    \frac{dI}{dt}(r,z,t)=\frac{\delta I}{\delta t}_{0} + \frac{\delta I}{\delta t}_{eh}
\end{equation}

The first term in this equation is the focal volume considering no carrier generation and the second term is the modification to this term due to the carrier plasma creating a spatially and temporally varying refractive index.

The initial term is found through Fourier optics in conjunction with the slow varying envelope approximation considering an weighted superposition of plane waves \cite{Goodmann}. Due to the large refractive index of diamond, a significant spherical aberration occurs at the interface, however by applying the opposing phase to the SLM, this can be mitigated. We follow work by Jesacher \emph{et al.} \cite{Jesacher} to separate out focal aberrations and calculate an unaberrated focal volume with a focal depth at 20 $\mu$m. The time dependence of the electric field was modelled as a Gaussian multiplied by a sinusoid of the laser frequency, $\omega_0$. The width is defined such that the full width half maximum (FWHM) of the squared Gaussian envelope is equal to the pulse duration. The spatial and time dependence's are multiplied together and the propagation of the pulse is included by an additional phase factor due to the differing phase velocities across the focal region.

\small
\begin{equation} \tag{S2}
    I_0(x,y,z,t)=I_V(x,y,z) \cdot I_t\Bigg(t+\frac{\omega_0 n^2 (z_0-z)}{c^2 \phi(x,y)}\Bigg)
\end{equation}
\normalsize

Where $\phi(x,y)$ is the normalised phase contributing towards the focusing and defocusing of the beam. This can be simplified by making use of azimuthal symmetry into $I(r,z,t)$ and rotating through 2$\pi$ about $z$ gives the full distribution. Multiplying by a scaling factor allows simulation of laser pulses of different pulse energies.

The second term in equation S1 regards the changes to this distribution due to the absorption of photons by electrons within the material and the changes to how the pulse propagates due to the generated carrier plasma. This is non-trivial to calculate as several processes occur that affect this term, namely photoionisation, free carrier absorption, plasma resonance, beam attenuation, self focusing and filamentation \cite{Mao} \cite{Kolesik}. The generation of carrier plasma (outlined in the following subsection) creates a spatially and temporally varying complex refractive index leading to absorption and attenuation of the pulse, reflection at regions where the refractive index changes and variations in phase velocity across the focal region. The reflected wave interferes with the incoming light and alters the intensity distribution further. These effects are calculated by substituting the spatially and temporally varying complex refractive index into the homogeneous wave equation governing the pulse propagation.

Figure \ref{plasmaspace}b, c and d compare the spatial intensity distribution at the peak intensity when the effect of carriers is considered for 3 different pulse energies and figure \ref{plasmaspace}a shows the distribution when plasma effects are neglected. Figure b shows a 5 nJ pulse. A small attenuation can be seen just below the focal depth as the carrier generation creates a small imaginary component to the refractive index. In contrast, higher energy pulses (7 and 8 nJ) produce a significant enough density of carriers that the effect on the incoming beam is drastic and attenuation starts to dominate. The reflected wave also interferes with the incoming wave leading to a small variation just above focus and the increase in phase velocity at the absorbing region can clearly be seen to bend the rays. These figures are plotted across a full diameter rather than radius as it is clearer to see the shifts in phase and spatial dependence of the attenuation. Plasma effects are generally insignificant for peak focal intensities lower than $\sim$ 40 TWcm$^{-2}$ however for pulse energies where the no-plasma peak intensity is greater than this, the carrier generation leads to significant attenuation such that increasing the pulse energy does not significantly increase the true peak focal intensity. This leads in part to the saturation of carrier and defect growth with respect to pulse energy.

\renewcommand{\thefigure}{S2}
\begin{figure}
    \centering
    \includegraphics[width=8cm]{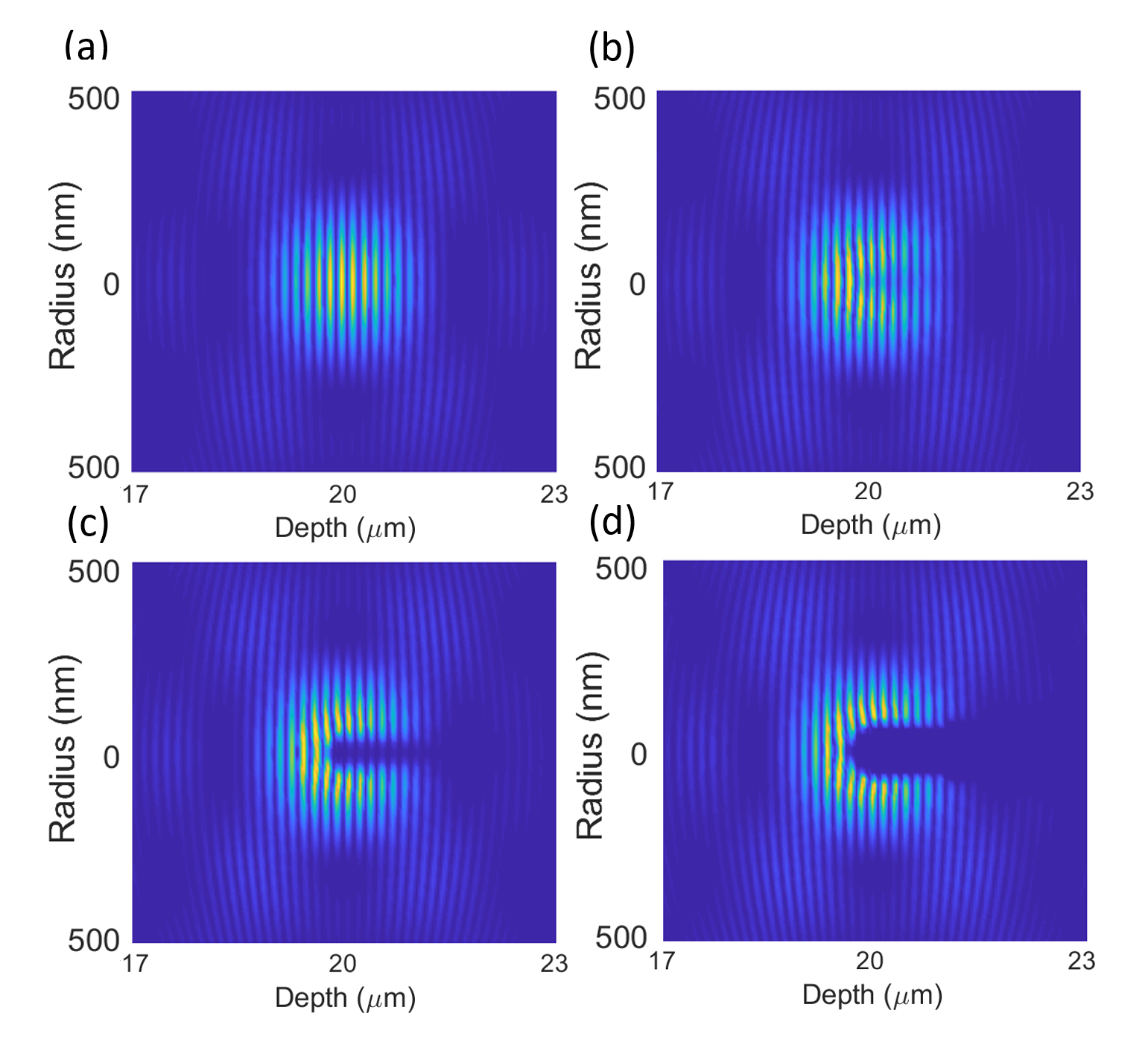}
    \caption{The spatial dependence of the light intensity at the peak time considering a. No plasma effects. b. A 5 nJ pulse, where peak intensities reach the threshold for attenuation to occur and the carrier generation has only a small effect. c. A 7 nJ pulse where plasma effects start to dominate d. A high energy (8 nJ) pulse where carrier generation leads to significant attenuation of the beam.}
    \label{plasmaspace}
\end{figure}

\subsection{Carrier Density}

The second rate equation concerns the density of electrons within the conduction band as a function of space, time and energy.

\small
\begin{equation} \tag{S3}
\begin{split}
    \frac{d\rho_{e}}{dt}(r,z,t,\mathcal{E}) = \frac{\delta \rho_e}{\delta t}_{PI} + \frac{\delta \rho_e}{\delta t}_{epp} + \frac{\delta \rho_e}{\delta t}_{e-ph} + \frac{\delta \rho_e}{\delta t}_{cc}\\ + \frac{\delta \rho_e}{\delta t}_{Casc} + \frac{\delta \rho_e}{\delta t}_{Auger} + \frac{\delta \rho_e}{\delta t}_{Rad}\\  + \frac{\delta \rho_e}{\delta t}_{Diff} + \frac{\delta \rho_e}{\delta t}_{Exc} + \frac{\delta \rho_e}{\delta t}_{Exc_{Diss}}
\end{split}
\end{equation}
\normalsize

The first term, $PI$ corresponds to the photoionisation of electrons across the band-gap and the following three terms refer to energy change processes through collisions; free carrier absorption via electron-phonon-photon collisions $epp$, electron-phonon scattering $ep$ and carrier-carrier scattering $cc$. The $Casc$ term refers to cascade or avalanche multiplication of the carrier density and the $Auger$ and $Rad$ terms regard losses to the electron density through Auger and radiative recombination processes. The $Diff$ term regards diffusion of electrons and the latter two terms, $Exc$ and $Exc_{Diss}$ regard the formation and dissociation of excitons.

Through charge neutrality, the density of holes generated through photoionisation is equal to the electron generation such that an identical equation exists for holes as for electrons although the rates of various scattering terms are different. 

The large band-gap of diamond means that single photons do not have enough energy to promote electrons across the potential barrier; photoionisation must therefore occur through non-linear processes involving the simultaneous absorption of multiple photons and tunnelling ionisation. This process is often modelled for many materials using the Keldysh equation which assumes a direct gap and linear dispersion \cite{Keldysh}. However curvature of the bands leads to differing photoionisation such that for materials with differing dispersion behaviour, the Keldysh approach is inaccurate \cite{Gruzdev}. A Monte Carlo approach to photoionisation for the diamond band-structure reveals that degeneracy in the valence bands at the gamma point leads to a nonperturbative quantum interference causing higher harmonic generation within the polarisation density of the crystal, significantly increasing the photoionisation rate at momentum states away from the centre of the Brillioun zone \cite{Lagomarsino}. Here we find that by making use of points of high symmetry, a hybrid of the Monte Carlo approach and the Keldysh approach leads to an accurate representation of photoionisation but with much less computational expense than a full Monte Carlo simulation.

The band structure of diamond was calculated using Density Function Theory (DFT) via CASTEP \cite{SJClarke}. Simulations were of a 3$\times$3$\times$3 supercell of diamond, in order to reduce the effects of the periodic boundary conditions, and employed the PBE exhange-correlation functional \cite{Perdew}. The simulation used the formation energy of the NV\textsuperscript{0} centre as a convergence parameter to determine an appropriate Monkhorst-Pack grid spacing and plane wave basis size. Following this, the band-gap was corrected such that the in-direct gap was equal to 5.41 eV. The closest bands to the bandgap are shown figure S3a.

In the long-wavelength approximation the canonical momentum gained by an electron within the field is given by the magnitude of the vector potential multiplied by the charge of the electron \cite{Polo}. This corresponds to an adiabatic horizontal shift of the valence bands in k space whilst leaving the conduction bands stationary allowing for photoionisation to take place away from the $\Gamma$ point. We make the assumption that for a small enough region of momentum space, and given an empirical ionisation potential, Keldysh theory should give a good approximation. Therefore we horizontally shift the valence bands by the appropriate momentum, and then utilise conventional Keldysh theory replacing the effective ionisation potential with the energy gap between the maximum point of the shifted bands to the unshifted conduction bands at the same momentum state. We define the polarisation density empirically to include the non-perturbative higher harmonic generation and calculate an effective instantaneous intensity felt by an electron including this harmonic generation, then replace the variable for intensity in the Keldysh equations with the effective intensity. Whilst this approach neglects some possible transitions, it captures most as the probability for transition is greatest at the point corresponding to the shifted gamma point due to the high density of states from high curvature of the bands. Iterating this for the entire duration of the pulse reveals close agreement with the Monte Carlo approach as can be see by comparing figure S3b with figure 11 of ref \cite{Lagomarsino}.

\renewcommand{\thefigure}{S3}
\begin{figure}
\centering
\includegraphics[width=8cm]{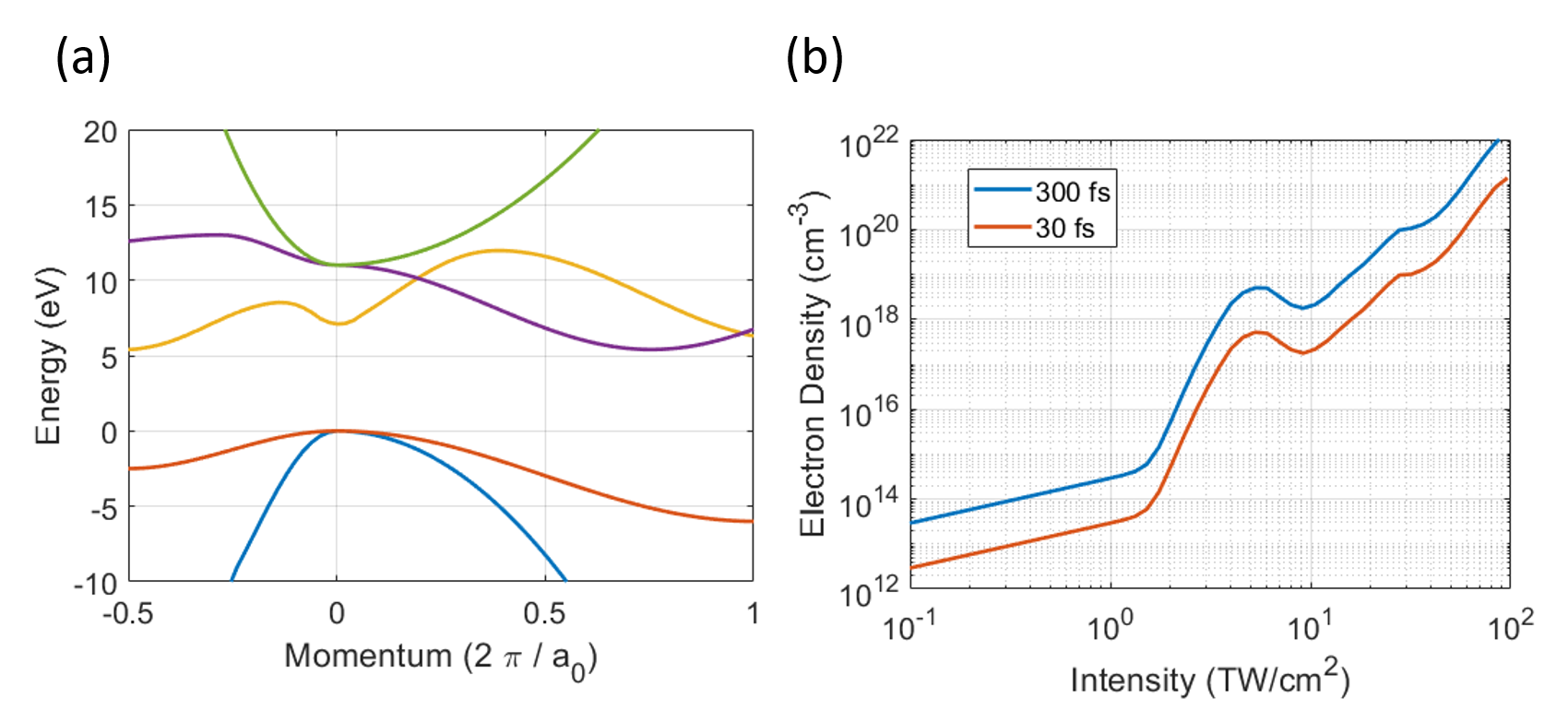}
\caption{a. The band-structure of diamond calculated using Density Function Theory (DFT) via CASTEP. b. The electron distribution generated via photo-ionisation as a function of laser intensity after a laser pulse of 300 fs and 30 fs.}
\label{fig:figs1}
\end{figure}

The energy distribution of conduction band states due to photoionisation correspond to near delta functions at the energy corresponding to the band locations at the momenta at which transition occurred. Many scattering events then take place which change the carrier distribution towards and away from equilibrium. Elastic scattering leads to momentum relaxation and will distribute the carriers more evenly across the Brillouin zone and inelastic scattering leads to energy relaxation and will ultimately bring the carriers back towards equilibrium with the lattice \cite{Boer}. Given no driving field, collisional energy losses average out such that the distribution thermalises through screened Coulomb interaction carrier-carrier collisions to an effective temperature much hotter than the lattice. Electron-phonon coupling causes the phonon subsystem to react to the non-equilibrium distribution of electrons generating longitudinal optical and acoustic phonons which are emitted from carriers transferring energy to the lattice \cite{Tandon}. The carrier-phonon coupling strength in the conduction band is much stronger than in the valence band, therefore the relaxation of holes takes place over a longer timescale \cite{Monserrat}.  These scattering events lead to cooling of the carrier density to the Fermi-Dirac distribution in equilibrium with the lattice. Under a driving electric field, carriers can absorb photons if their oscillation is knocked out of phase with the incoming field \cite{Guenther}. This is known as reverse Bremsstrahlung or carrier-phonon-photon scattering which leads to an efficient energy gain of carriers, skewing the distribution away from equilibrium. The photoionisation and reverse Bremsstrahlung processes compete with the phonon and carrier scattering leading to a skewed thermal distribution. 
We approximate the rates of each of these terms as a function of space, time and energy by using a mixture of collision time approximations, Drude theory and energy and momentum conservation \cite{Mao} \cite{Shen}. The average carrier-carrier collision time is calculated using the change to the carrier mobility as a function of carrier density under the assumption that each scattering event randomises the motion \cite{Kozak} given by the equation.

\small
\begin{equation} \tag{S4}
\begin{split}
    \tau_{cc}(r,z,t)=\frac{m' a T(r,z,t)^\frac{2}{3}}{e N_{carr}(r,z,t)}\\\times\frac{1}{\ln(1 + b T(r,z,t)^2 (1\times10^{-4} \cdot N_{carr}(r,z,t))^{-\frac{2}{3}})}
    \end{split}
\end{equation}
\normalsize

Where $N_{carr}$ is the total carrier density inclusive of electrons, holes and their bound states. The density dependent nature of this term means that it is negligible at low densities with timescales of order tens of pico-seconds for carrier densities of 10$^{17}$ cm$^{-3}$ but dominates at high carrier densities where the timescale is reduced to less than a femtosecond for carrier densities of $\sim$ 10$^{21}$ cm$^{-3}$ vastly reducing mobility. Carrier scattering is elastic; the total energy is conserved but the scattering acts to homogenise the momentum distribution across the Brillouin zone and leads to thermalisation towards a Boltzmann distribution. This effect can be calculated considering carrier scattering between two carriers with the constraint that the total energy must be conserved and integrating across energy. This is complicated slightly by the density of states which limits the states which electrons can scatter into at high densities. The change in the electron distribution due to carrier-carrier scattering is given by

\small
\small
\begin{equation} \tag{S5}
\begin{split}
    \frac{\delta \rho_e}{\delta t}_{cc} (r,z,t,\mathcal{E})=
   \frac{1}{\tau_{cc}(r,z,t)} \Bigg[ \int_{\mathcal{E}_C}^{\infty}\int_{\mathcal{E}_C}^{\infty} \frac{\rho_e(r,z,t,\mathcal{E}_1)}{\mathcal{E}_1+\mathcal{E}_2 - \mathcal{E}_C} \\ \times
    \frac{\rho_e(r,z,t,\mathcal{E}_2)}{N_{carr}(r,z,t)}\ d\mathcal{E}_1 \  d\mathcal{E}_2 - \rho_e(r,z,t,\mathcal{E}) \Bigg] \ \\ \text{for} \ \mathcal{E} \le \mathcal{E}_1+\mathcal{E}_2 - \mathcal{E}_C
\end{split}
\end{equation}
\normalsize

Under the condition that

\begin{equation} \tag{S6}
 \rho_e(r,z,t,\mathcal{E})+\delta \rho_{e_{cc}} (r,z,t,\mathcal{E}) < D(\mathcal{E})
\end{equation}

Where $D(\mathcal{E})$ is the density of states at a given energy. This heads towards a Boltzmann distribution given many scattering events. We find the density of states by numerically integrating the momenta states from the band structure over energy iso-surfaces.

Scattering with acoustic phonons is a near elastic process whereas optical phonons scatter inelastically with carriers and dominate energy transfer \cite{Boer}. Tandon \emph{et al.} used density functional theory to calculate the average carrier-phonon scattering rate as a function of energy inclusive of emission and absorption within the valence and conduction bands \cite{Tandon}. They found that the phonon scattering rate is proportional to the density of electron states at a particular energy.

\begin{equation} \tag{S7}
    \tau_{ep_{tot}} (\mathcal{E})=\frac{1}{a D(\mathcal{E})}
\end{equation}

With a proportionality constant of $a=1\times 10^{-12}$ s$^{-1}$m$^{-3}$. Monserrat \emph{et al.} used density functional theory to calculate the phonon energies  \cite{Monserrat} for which we average to find an optical phonon energy of 0.15 eV and 0.08 eV for acoustic phonons. We determine the ratio of phonon emission to absorption by comparing the lattice temperature to a calculated electron temperature where we calculate the electron temperature by fitting a best fit exponential function to the electron energy distribution. This then allows for calculation of phonon emission and phonon absorption rates given by

\small
\begin{equation} \tag{S8}
  \tau_{ep_{emit}}(r,z,t,\mathcal{E})= \tau_{ep_{tot}}(\mathcal{E})\frac{T_{e}(r,z,t)+T_{L}(r,z,t)}{T_{e}(r,z,t)}
\end{equation}

\begin{equation} \tag{S9}
  \tau_{ep_{abs}}(r,z,t,\mathcal{E})= \tau_{ep_{tot}}(\mathcal{E})\frac{T_{e}(r,z,t)+T_{L}(r,z,t)}{T_{L}(r,z,t)}
\end{equation}
\normalsize

A proportion of the carriers at a given energy then moves to a higher or lower energy dependent upon whether the phonon is absorbed or emitted.

\small
\begin{equation} \tag{S10}
\begin{split}
    \frac{\delta \rho_e}{\delta t}_{ep}(r,z,t,\mathcal{E}) = -\frac{\rho_e(r,z,t,\mathcal{E})}{\tau_{ep_{tot}}(r,z,t,\mathcal{E})}\\ + \frac{\rho_e(r,z,t,\mathcal{E} - \mathcal{E}_{ph})}{\tau_{ep_{abs}}(r,z,t,\mathcal{E} - \mathcal{E}_{ph})} + \frac{\rho_e(r,z,t,\mathcal{E} + \mathcal{E}_{ph})}{\tau_{ep_{emit}}(r,z,t,\mathcal{E} + \mathcal{E}_{ph})}
\end{split}
\end{equation}
\normalsize

This approach neglects higher order terms where two or more phonons interact at the same time but approaches the true result given many scattering events. The average rate of energy gained by carrier through electron-phonon-photon (reverse Bremsstrahlung) processes can be calculated through Drude theory inputting a combination of the carrier-phonon and carrier-carrier scattering rates added together through Matthiessen's rule to form $\tau_{Scatt}$.

\small
\begin{equation} \tag{S11}
    \frac{\delta \mathcal{E}}{\delta t}_{epp} (r,z,t,\mathcal{E}) = \frac{(\frac{e^2 I(r,z,t)}{m' c \epsilon_0 n (r,z,t,\mathcal{E})})\tau_{scatt}(r,z,t,\mathcal{E})}{\omega_0^2 \tau_{scatt}(r,z,t,\mathcal{E})^2 + 1} 
\end{equation}
\normalsize

Through similar arguments to above, this can be converted into the change to the carrier density  as a function of energy by considering that carriers can only gain energy in discrete amounts corresponding to the energy of a photon or the energy of a photon plus or minus the energy of a phonon. 

Carriers above a threshold energy, $\mathcal{E}_{th}$ can relax to lower energy states through impact ionisation leading to the generation of an electron-hole pair. The reverse process of this only becomes important at very high carrier densities and causes a recombination of an electron-hole pair through the Auger effect with the excess energy being passed to a third carrier. We calculate the rates of cascade and Auger processes through density and energy conservation arguments across both bands considering a 3 carrier collision for carriers above a threshold energy. The cascade term is given by

\small
\begin{equation} \tag{S12}
\begin{split}
    \frac{\delta \rho_e}{\delta t}_{casc}(r,z,t,\mathcal{E}) = \int_{\mathcal{E}_{th}}^{\infty} \int_{\mathcal{E}_{th}-\mathcal{E}_2}^{0}
    \frac{\rho_e(r,z,t,\mathcal{E})}{\tau_{cc}(r,z,t)}\\ \bigg(\frac{D(\mathcal{E}_1)-\rho_h(r,z,t,\mathcal{E}_1)}{D(\mathcal{E}_1)}\bigg)\\\bigg(\frac{D(\mathcal{E}_2)-\rho_e(r,z,t,\mathcal{E}_2)}{D(\mathcal{E}_2)}\bigg) \ d\mathcal{E}_1 \ d\mathcal{E}_2
\end{split}
\end{equation}
\normalsize

Where and $\mathcal{E}_1$ is the initial hole energy and $\mathcal{E}_2$ is the initial energy of the high energy electron. The hole energy is negative as we define the zero point of energy as the maxima of the valence band. The terms in the first and second bracket are approximately 1 until the density of carriers becomes very high, hence, once integrated across energy this simplifies to the cascade rate being proportional to the density of carriers for a low carrier density. Auger processes, being the reverse of cascade, are governed by a very similar equation, by considering the density of occupied states rather than free states. When integrated across energy this is proportional to the cube of the carrier density.  

Radiative recombination is purely density dependent with a constraint of no shift in momenta and is regulated through the screened Coulomb interaction \cite{Lipatov}. Diffusion of carriers within the material will be influenced by local strain fields, however to first approximation we assume strain is homogeneous and utilise a diffusion equation purely based upon density and mobility determined by the Einstein relation. The free carrier density can also be reduced through the formation of excitonic states by electrons coupling to holes through a screened Coulomb attraction.

Strictly speaking electron temperature only makes physical sense when the system is in equilibrium. In this highly dynamical system, the system is far from equilibrium however, if we plot the energy dependence of the carrier density on a logarithmic-linear scale, equilibrium carrier distributions are given by straight lines and deviation from this represents the non-equilibrium dynamics.

\renewcommand{\thefigure}{S4}
\begin{figure}
    \centering
    \includegraphics[width=7cm]{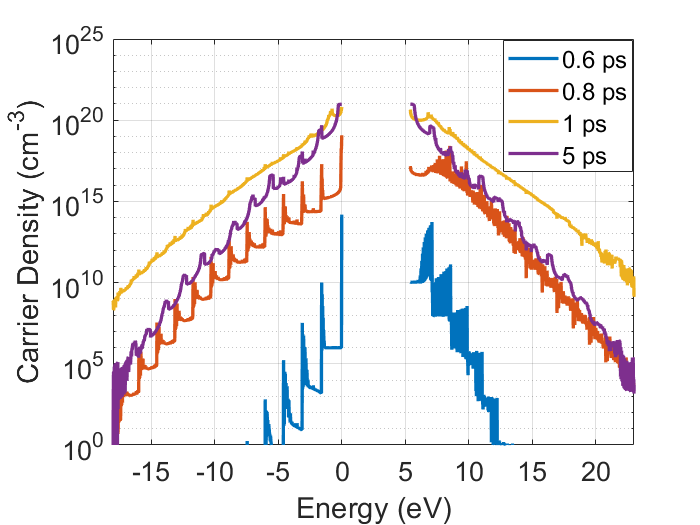}
    \caption{An example of carrier energy distribution at the centre of focus at various times throughout the pulse where the zero of energy is the maxima of the valence band. Carriers of energy below zero are holes and carriers above zeros are electrons.}
    \label{fig:S2}
\end{figure}

Initially photoionisation produces sharp deviations from equilibrium throughout. Scattering processes quickly thermalise the distribution such that an effective carrier temperature can be calculated. We see that the carrier temperature grows quickly to thousands of degrees and that lattice temperature remains low for a much longer period. Following the pulse, carrier temperatures decay over tens of picoseconds and thermalise with the lattice.

\subsection{Excitonic States}

An electron and hole with overlapping wavefunctions can become bound through the screened Coulomb interaction and form excitonic states if their kinetic energy is less than the excitonic binding energy \cite{Mao} \cite{Selbmann}.  We calculate the exciton formation rate via

\begin{equation} \tag{S13}
\begin{split}
    \frac{\delta \rho_{Exc}}{\delta t}_{Form}(r,z,t,\mathcal{E})=\\ \int_{\mathcal{E}_{C}}^{\mathcal{E}}  \frac{\rho_e(r,z,t,\mathcal{E}_{1})\rho_h(r,z,t,\mathcal{E}_{1}-\mathcal{E}-\mathcal{E}_{x})}{N_{carr}(r,z,t) \cdot \tau_{cc}(r,z,t)} \ d\mathcal{E}_{1}
\end{split}
\end{equation}

Where $\mathcal{E}_{X}$ is the excitonic binding energy \cite{Clark2} and $\mathcal{E}_{C}$ refers to the conduction band minimum and is the maximum possible energy of the bound pair. This rate is doubly dependent upon the density of free carriers as the interaction timescale is also density dependent. Single excitons in diamond are of Wannier type and remain relatively spread out over space \cite{Mauri}, therefore excitonic binding is weak and the electron-hole pair continue to scatter phonons and other carriers somewhat independently. We keep track of the density of excitons as a function of space, time and energy in a further coupled equation.

\begin{equation} \tag{S14}
\begin{split}
    \frac{d\rho_{Exc}}{dt}(r,z,t,\mathcal{E})= \frac{\delta \rho_{Exc}}{\delta t}_{Form}  + \frac{\delta \rho_{Exc}}{\delta t}_{Exc-c}\\ + \frac{\delta \rho_{Exc}}{\delta t}_{Exc-ph} + \frac{\delta \rho_{Exc}}{\delta t}_{Recomb} + \frac{\delta \rho_{Exc}}{\delta t}_{BiX}\\ + \frac{\rho_{Exc}}{\delta t}_{BiX_{Diss}} + \frac{\delta \rho_{Exc}}{\delta t}_{BiX_{Recomb}}
\end{split}
\end{equation}

The first term corresponds to the formation of excitons, the second term corresponds to scattering processes with carriers which are similar to those described above for free carriers and thermalise the distribution also leading to dissociation of excitons back into free carriers when the exciton absorbs energy greater than the binding energy. During the pulse and shortly after the pulse, the incoming light means that excitons which form are quickly dissociated, however once the carriers and lattice form a quasi-equilibrium, most interactions lead to a loss of energy rather than a gain such that the carriers remain bound. The third term corresponds to scattering with phonons and the fourth term to excitonic recombination which takes place either radiatively, through phonon scattering, Auger processes or existing defect trapping on timescales of nanoseconds \cite{Morimoto}. The $BiX$ term corresponds to the combining of excitonic states into biexcitons, again through the screened Coulomb interaction and the final two terms regard the gain to the exciton density from the dissociation and recombination of biexcitons. As before, we define a coupled equation to keep track of the density of biexcitons as a function of space, time and energy with interactions derived similarly.

\small
\begin{equation} \tag{S15}
\begin{split}
    \frac{d\rho_{BiX}}{dt}(r,z,t,\mathcal{E})= \frac{\delta \rho_{BiX}}{\delta t}_{Form} + \frac{\delta \rho_{BiX}}{\delta t}_{BiX-c}\\ + \frac{\delta \rho_{BiX}}{\delta t}_{BiX-ph} + \frac{\delta \rho_{BiX}}{\delta t}_{Rad}
    \end{split}
\end{equation}
\normalsize

The rate at which biexcitons form is given by

\small
\begin{equation} \tag{S16}
\begin{split}
    \frac{\delta \rho_{BiX}}{\delta t}_{Form}(r,z,t,\mathcal{E})= \\ \int_{\mathcal{E}_{C}-\mathcal{E}_{x}}^{\mathcal{E}} \frac{\rho_{Exc}(r,z,t,\mathcal{E}_{1})\rho_{Exc}(r,z,t,\mathcal{E}-\mathcal{E}_{1}+\mathcal{E}_{bx})}{N_{carr}(r,z,t)\cdot \tau_{xx}(r,z,t)} \ d\mathcal{E}_{1} 
\end{split}
\end{equation}
\normalsize

Where $\mathcal{E}_{bx}$ is the biexcitonic binding energy which is a further 12 meV for diamond \cite{Katow} and $\tau_{xx}$ is the equivalent of the carrier-carrier scattering time but just considering excitonic density. The scattering terms in equation S16 include dissociation into single excitons and thermalisation. The fourth term in the biexciton equation regards radiative recombination, the rate of which we determine from \cite{Omachi}. There should be a fifth term in this equation along with another coupled equation corresponding to the formation of higher order multi-excitons but for simplicity we neglect these terms.
Biexcitons are more bound than single excitons such that the lowest energy configuration of the lattice-biexciton system corresponds to a significantly deformed lattice and the biexciton trapped at a single atomic site. Upon emission of phonons, biexcitons will self-trap, allowing relaxation of the lattice, releasing 1.74 eV and breaking a carbon-carbon bond  \cite{Mauri}.  We keep track of the density of self-trapped biexcitonic states in a further equation

\small
\begin{equation} \tag{S17}
\begin{split}
    \frac{d\rho_{STbX}}{dt}(r,z,t,\mathcal{E})= \frac{\delta \rho_{STbX}}{\delta t}_{Form} + \frac{\delta \rho_{STbX}}{\delta t}_{STbX-c}\\ + \frac{\delta \rho_{STbX}}{\delta t}_{ph} + \frac{\delta \rho_{STbX}}{\delta t}_{Rad} + \frac{\delta \rho_{STbX}}{\delta t}_{FD}
  \end{split}
\end{equation}
\normalsize
The final term describes the hypothesised non-radiative recombination yielding the formation of a Frenkel defect. The higher localisation and higher energies of self-trapped multi-excitonic states mean that these act as a source of energy localisation within the crystal.

\subsection{Frenkel Defects}

Self-trapped multi-excitonic states allow for energy localisation and quasi-particles of high energies whilst simultaneously breaking a carbon-carbon bond through a deformation potential of $\mathcal{E}_{DP}=$1.74 eV \cite{Mauri}. Outlined in the main text we propose a hypothesis for the generation of Frenkel pairs by the non-radiative recombination of self-trapped multi-excitonic states. We include another coupled equation corresponding to the hypothesised density of lattice defects given by

\begin{equation} \tag{S18}
\begin{split}
    \frac{d\rho_{FD}}{dt}(r,z,t)= \\
    \int_{2(\mathcal{E}_{C}-\mathcal{E}_{X})-\mathcal{E}_{bX}-\mathcal{E}_{DP}}^{2(\mathcal{E}_{C}-\mathcal{E}_{X})-\mathcal{E}_{DP}} \frac{\rho_{STbX}(r,z,t,\mathcal{E})}{\tau_{FD}(r,z,t,\mathcal{E})}\ d \mathcal{E}
\end{split}
\end{equation}

\subsection{Lattice Temperature Distribution}

The lattice temperature is calculated according to the following equation

\begin{equation} \tag{S19}
    \frac{d T_L}{dt}(r,z,t) = \frac{\delta T_L}{\delta t}_{Scatt} + \frac{\delta T_L}{\delta t}_{Diff}
\end{equation}

The first term regards lattice heating through the phonon scattering of carriers and excitons and is calculated by adding up the energy lost from the various scattering centres. For example, the temperature gain from phonon emission from electrons is given by

\begin{equation} \tag{S20}
    \frac{\delta T_L}{\delta t}_{Scatt_{e}} =  \int_{0}^{\infty}  \frac{\mathcal{E}_{ph}}{k_B}\frac{\rho_e(r,z,t,\mathcal{E})}{\tau_{ep_{emit}}(r,z,t,\mathcal{E})\rho_A} \ d\mathcal{E}
\end{equation}

Where $\rho_A$ is the atomic density. Similar terms exist for the other scattering terms. These scattering terms act to bring the lattice carrier system to a quasi-equilibrium state. The second term describes the diffusion of heat throughout the lattice for which we use a simple heat equation model.

\subsection{Worthy of Note}

The spatial distribution of defects that are generated becomes skewed towards shallower depths as the pulse energy increases due to the photogenerated plasma obtaining a metallic phase and leading to significant absorption and attenuation of the incoming laser beam. An example of this can be seen in figure \ref{fig:S3} where the spatial distribution of Frenkel defects generated by a 7 nJ pulse is shown.

\renewcommand{\thefigure}{S5}
\begin{figure}
    \centering
    \includegraphics[width=8cm]{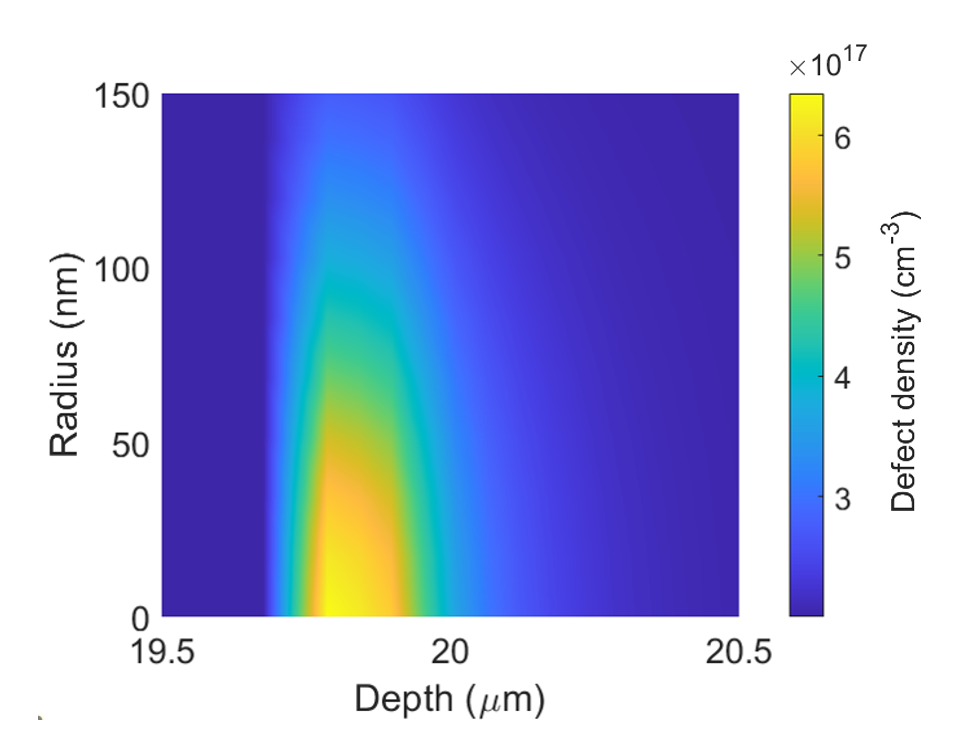}
    \caption{The spatial distribution of Frenkel defects generated following a 7 nJ pulse. The distribution is skewed to shallower depths as plasma is generated during the focusing of the pulse and leads to attenuation of the beam.}
    \label{fig:S3}
\end{figure}

The rates of many terms within the coupled equations depend upon empirical scattering rates taken from the literature shown in table I. These are a mixture of experimentally theoretically derived parameters and many contain significant error bars. The results of this work do not claim that the laser forms a specific number of vacancies, instead we note the trends, the effective non-linearity, determined given the hypothesis of multi-excitonic recombination.

Above a critical density, excitonic states undergo a Mott transition into an electron-hole liquid. In this analysis we neglect this transition as the critical density at the high temperatures simulated in this work is above the breakdown threshold for graphitisation. Due to the long life-times of electron hole pairs, the radiative recombination terms in the carrier equations can be neglected without a visible change to the results. Neglecting carrier diffusion appears to slightly reduce the pulse energy threshold for defect formation and breakdown as the high carrier densities are reached at a slightly earlier stage.

Yurgens \emph{et al.} recently showed that the use of a truncated hemispherical cubic zirconia solid immersion lens (t-SIL) could drastically reduce the threshold write-pulse energy for vacancy creation \cite{Yurgens}. They used single 35 fs, 800 nm pulses focused through a 0.9 NA air lens into the t-SIL and achieved vacancy creation at pulse energies down to 5.8 nJ. Without the use of the t-SIL, they were unable to generate vacancies up-to the pulse energy limit of their laser (52 nJ). The reason for this drastic reduction in pulse energy threshold is likely due to a higher spatial confinement of the laser focus by the t-SIL allowing for much higher focal intensities at low pulse energies than the traditional optical methods used in this work.

\end{document}